\documentclass[letterpaper, 10 pt, conference]{ieeeconf}  % Comment this line out if you need a4paper

\IEEEoverridecommandlockouts                              % This command is only needed if 
\usepackage{amsmath}   % equations, cases, align, \lVert...\rVert, etc.
\usepackage{amssymb}   % extra math symbols
\usepackage{amsfonts}  % \mathbb
\usepackage{mathtools} % small math tweaks, e.g., \DeclarePairedDelimiter

\usepackage{bm}

\usepackage{graphicx}

\usepackage[hidelinks,breaklinks=true]{hyperref}
\usepackage{xurl}
\usepackage[capitalise,nameinlink]{cleveref}
\usepackage{cite}
\usepackage{xcolor} 
\usepackage{comment}
\usepackage{subcaption} % da inserire nel preambolo
\usepackage{tikz}
\usetikzlibrary{shapes, arrows.meta, positioning, calc}

\title{\LARGE \bf

Decentralized Shepherding of Non-Cohesive Swarms Through Cluttered Environments via Deep Reinforcement Learning}

\author{Cristiana Punzo$^{1}$, Italo Napolitano$^{1,\dag}$, Cinzia Tomaselli$^{1,\dag}$, Mario di Bernardo$^{1,2 *}$
\thanks{The authors acknowledge support from the Italian Ministry of University and Research (MUR) under project PRIN 2022 ``Machine-learning based control of complex multi-agent systems for search and rescue operations in natural disasters (MENTOR).''}
\thanks{$^{1}$Cristiana Punzo, Italo Napolitano, Cinzia Tomaselli and Mario di Bernardo are with the Modeling and Engineering Risk and Complexity Department, Scuola Superiore Meridionale, via Mezzocannone 4, 80138, Naples, Italy (email: cr.punzo@studenti.unina.it, i.napolitano@ssmeridionale.it, c.tomaselli@ssmeridionale.it, mario.dibernardo@unina.it)}%
\thanks{$^{2}$Mario di Bernardo is with the Department of Information Technology and Electrical Engineering, University of Naples Federico II, Naples, Italy.}%
\thanks{\textsuperscript{\dag} These authors contributed equally to this work. }
\thanks{* Corresponding author}
}

\usepackage{graphicx}
\usepackage[percent]{overpic}

\begin{document}

\maketitle
\thispagestyle{empty}
\pagestyle{empty}

%%%%%%%%%%%%%%%%%%%%%%%%%%%%%%%%%%%%%%%%%%%%%%%%%%%%%%%%%%%%%%%%%%%%%%%%%%%%%%%%
\begin{abstract}

This paper investigates decentralized shepherding in cluttered environments, where a limited number of herders must guide a larger group of non-cohesive, diffusive targets toward a goal region in the presence of static obstacles. A hierarchical control architecture is proposed, integrating a high-level target assignment rule, where each herder is paired with a selected target, with a learning-based low-level driving module that enables effective steering of the assigned target. The low-level policy is trained in a one-herder–one-target scenario with a rectangular obstacle using Proximal Policy Optimization and then directly extended to multi-agent settings with multiple obstacles without requiring retraining. Numerical simulations demonstrate smooth, collision-free trajectories and consistent convergence to the goal region, highlighting the potential of reinforcement learning for scalable, model-free shepherding in complex environments.

\end{abstract}

%%%%%%%%%%%%%%%%%%%%%%%%%%%%%%%%%%%%%%%%%%%%%%%%%%%%%%%%%%%%%%%%%%%%%%%%%%%%%%%%
\section{Introduction}
The shepherding control problem, where a small number of controlled agents (herders) must guide a larger group of diffusive agents (targets) toward a goal region, is fundamental to applications ranging from emergency evacuation and oil spill containment to drone defense and search-and-rescue operations~\cite{licitra, long, pierson, liu, aerialdefense}. Inspired by natural phenomena like sheepdogs herding livestock through gates and around barriers, this problem provides a compelling framework for studying indirect control in heterogeneous multi-agent systems where coordination must be achieved without centralized communication or explicit cooperation from the targets.

Classical shepherding formulations assume targets exhibit cohesive flocking behavior through local attraction, repulsion, and alignment forces, enabling herders to treat the group as a single controllable entity~\cite{strombom2014, zheng2024, jadhav2024}. These approaches have provided valuable theoretical insights and work well in controlled settings. However, real-world scenarios systematically violate this cohesion assumption. Evacuating crowds scatter in panic rather than maintaining group coherence, contaminated particles diffuse independently following stochastic dynamics, and hostile drones actively evade capture using unpredictable maneuvers. Moreover, real environments contain obstacles such as walls, debris fields, and restricted zones that not only constrain motion paths but also occlude sensors, fragment target groups into disconnected subpopulations, and force indirect routing strategies that must account for future target dispersal. Current methods fail when confronting these dual challenges of non-cohesive targets and environmental constraints, severely limiting their deployment in the safety-critical applications where autonomous shepherding capabilities are most urgently needed.

Recent advances have begun addressing non-cohesive targets using hierarchical control architectures, where high-level controllers assign specific targets to individual herders while low-level controllers determine the optimal steering actions~\cite{auletta2022, lama2024, napolitano2024emergent, napolitano2025hierarchical, covone_hierarchical_2025, de2021application, nino}. These methods successfully demonstrate that non-cohesive populations can be controlled through careful coordination of individual influences. Yet they remain confined to idealized obstacle-free environments, leaving unresolved how such strategies would perform in cluttered spaces. 

Meanwhile, obstacle-aware shepherding methods have been developed to enable navigation around barriers, leveraging techniques such as geometric projections, repulsive potential fields, and control barrier functions~\cite{chipade, hamandi, zhang2024distributed}. Control strategies based on repulsive potential fields applied to both target and herder dynamics offer lightweight, fully local, and reactive control, but are susceptible to deadlocks and local minima. Conversely, approaches combining geometric projections and control barrier functions are generally integrated with global path-planning modules. As in other methods primarily relying on global planning~\cite{liu2023effective}, this integration introduces centralized computation requirements. To the best of our knowledge, all existing approaches assume cohesive target behavior--a condition rarely satisfied in realistic or emergency scenarios.

Learning-based approaches offer a promising alternative by discovering effective policies directly from trial-and-error experiences, potentially capturing complex behaviors that elude analytical modeling. While Deep Q-Networks have successfully learned to navigate cohesive flocks through structured obstacle fields~\cite{zhi2021learning, zhi2022}, demonstrating that neural policies can implicitly encode sophisticated spatial reasoning, the challenging problem of multi-agent shepherding with non-cohesive targets in cluttered environments remains largely unexplored.

We address this gap through a hierarchical architecture that combines PPO-based reinforcement learning for low-level herding behaviors with decentralized target assignment for multi-agent coordination. Our key insight is that obstacle-aware shepherding behaviors, learned in minimal scenarios with a single herder, target, and obstacle, can successfully transfer to complex multi-agent settings without retraining. This approach requires only minimal system modeling, thereby eliminating the need for online mapping and path planning, which are often centralized, computationally intensive, or suboptimal.
Numerical experiments validate the strategy and demonstrate superior gathering efficiency compared to traditional rule-based methods, establishing that deep reinforcement learning provides a scalable, model-free foundation for real-world shepherding challenges in complex domains.

\section{Problem statement}

\subsection{Modeling}
\paragraph{Environment geometry}
We consider a planar environment $\mathbb{R}^2$ in which $N$ herders must steer $M \geq N$ targets into a circular goal region $\Omega_\mathrm{G} \subset \mathbb{R}^2$, centered at the origin and with radius $\rho_\mathrm{G}$. 

The environment contains a set of \( C \) static rectangular  obstacles with smoothed corners, denoted \( \{ P_k \}_{k=1}^C \). 
We assume that each obstacle \( P_k \subset \mathbb{R}^2 \) is centered at \( \mathbf{P}_k \in \mathbb{R}^2 \), has fixed side lengths \( (L,S) \), and is oriented such that its long side is orthogonal to the line connecting \( \mathbf{P}_k \) to the center of \( \Omega_\mathrm{G} \). This configuration represents a worst-case scenario for the task, while allowing the obstacle’s area to be fully determined solely from the position $\mathbf{P}_k$.

Rectangular obstacles make the task inherently more challenging, as their flat sides tend to trap agents along the edges when only local normal repulsion is applied. Consequently, agents must actively learn obstacle-aware trajectories to avoid stagnation. This choice provides a minimal geometric configuration that naturally induces complex interactions, thereby increasing task difficulty with limited geometric complexity. The proposed framework can be extended to other obstacle geometries by retraining the policy accordingly.

To ensure geometric feasibility, obstacles are required to be non-overlapping and sufficiently spaced from one another. 
In particular, we assume the distance between the centers of any two rectangular obstacles satisfies the property: 

\begin{equation}
    \|\mathbf{P}_i - \mathbf{P}_j\|_2 \;>\; 
    \sqrt{(S + 2d_\mathrm{o}^*)^2 + (L + 2d_\mathrm{o}^*)^2}, 
    \quad \forall\, i \neq j,
\end{equation}
where the parameter $d_\mathrm{o}^*>0$ is a safety distance.

This condition enforces a minimum spacing between obstacles, ensuring that their rectangular bounding boxes, enlarged by a safety margin \( d_\mathrm{o}^* \), do not intersect. 
In particular, each obstacle is surrounded by an expanded buffer region of thickness \( d_\mathrm{o}^* \) on all sides; the Euclidean distance between the centers of any two obstacles must therefore exceed the diagonal length of these buffered boxes. 
This prevents overlapping or excessively close configurations that could create unrealistic narrow passages for the agents.

The initial conditions of all agents are constrained within a circular domain $\Omega_0 \subset \mathbb{R}^2$ of radius $R$ centered at the origin, excluding the obstacle regions, i.e.,
\begin{equation}
    \Omega_0 := \{\mathbf{x} \in \mathbb{R}^2 : \lVert\mathbf{x} \rVert_2 \leq R\; \land \; \mathbf{x} \notin \bigcup_{k=1}^C P_k \}.
\end{equation}

\paragraph{Targets' dynamics}
Targets evolve under overdamped stochastic dynamics with only short-range repulsion from herders and obstacles. Cohesion, alignment, and communication are excluded, so the population behaves as independent agents. This reflects dispersed or uncooperative settings and avoids artificial flocking, providing a conservative benchmark for evaluating control.
The target population is modeled with first-order differential equations under kinematic assumptions, considering a negligible acceleration phase duration \cite{albi2016invisible}.
Building on the minimal, cohesion-free model in~\cite{lama2024}, we augment the target dynamics with a short-range local repulsion from nearby obstacles, denoted $\mathbf{F}_{\mathrm{o}}(\mathbf{q})$.
We define the Cartesian coordinates of the target $T_i$ (for $i=1, \ldots, M$) as $\mathbf{T}_i(t)$ and stack them in the target population position vector $\mathbf{T}:=\left[\mathbf{T}_1, \dots, \mathbf{T}_M\right]$.
The complete dynamics of each target $i$ is:
\begin{equation}
\label{eq:target_dynamics_full}
\begin{aligned}
& \dot{\mathbf{T}}_i 
= \sqrt{2D}\,\boldsymbol{\mathcal{N}}
+ \left\{ \beta \sum_{j \in \mathcal{N}_{H,i}} \big(\lambda - \| \mathbf{d}_{ji} \|_2\big)\,\hat{\mathbf{d}}_{ji}
+ \mathbf{F}_\mathrm{o}(\mathbf{T}_i) \right\}_{v_T}, 
\end{aligned}
\end{equation}
where $\{ \cdot \}_a$ denotes an element-wise saturation operator ensuring $\| \cdot \|_\infty \leq  a$, \( \boldsymbol{\mathcal{N}} \) is a two-dimensional standard Gaussian white noise process with diffusion coefficient $D>0$, \( \beta>0 \) is the repulsion gain, and \( \lambda>0 \) is the herder–target interaction radius. The set of influencing herders is \( \mathcal{N}_{H,i}=\{\, j : \|\mathbf{T}_i-\mathbf{H}_j\|_2\le \lambda \,\} \). 
Denoting \( \mathbf{d}_{ji} = \mathbf{T}_i - \mathbf{H}_j \) and \( \hat{\mathbf{d}}_{ji} = \mathbf{d}_{ji}/\|\mathbf{d}_{ji}\|_2 \), the deterministic interaction term represents a linear (harmonic) radial repulsion that vanishes at distance \( \lambda \). 
The term \( \mathbf{F}_\mathrm{o}(\mathbf{q}) \) models local, short-range repulsion from nearby (rounded) rectangular boundaries.
Let \( \partial P_k \) be the boundary of obstacle \(k\) and
\begin{equation}
s_k(\mathbf{q}) \;=\; \min_{\mathbf{q}' \in \partial P_k} \| \mathbf{q} - \mathbf{q}' \|_2
\end{equation}
be the shortest distance of a point \( \mathbf{q} \in \mathbb{R}^2 \) from that boundary. The obstacle repulsion force is derived from compact-support artificial potentials~\cite{khatib1986real}, following the formulation in~\cite{zhi2022}:
\begin{subequations}\label{eq:Fc_def}
\begin{align}
\mathbf{F}_\mathrm{o}(\mathbf{q}) &= - \sum_{k=1}^{C} \nabla U_k(\mathbf{q}), \label{eq:Fc_def:force}\\
U_k(\mathbf{q}) &=
\begin{cases}
\dfrac{1}{2}\,k_\mathrm{o} \left( \dfrac{1}{s_k(\mathbf{q})} - \dfrac{1}{d_\mathrm{o}^\ast} \right)^{\!2}, & s_k(\mathbf{q}) \le d_\mathrm{o}^\ast,\\[6pt]
0, & s_k(\mathbf{q}) > d_\mathrm{o}^\ast,
\end{cases}\label{eq:Fc_def:Uk}
\end{align}
\end{subequations}
where \(k_\mathrm{o}>0\) is the interaction gain and \(d_\mathrm{o}^\ast>0\) is the activation distance. 
Importantly, the presence of this short-range potential with $d_\mathrm{o}^*$ relatively small (see Sec. \ref{sec:validation}) only prevents near-contact collisions and does not constitute a global obstacle-avoidance strategy.

\paragraph{Herders' dynamics}
We define the Cartesian coordinates of the herder $H_j$ (for $j = 1, \ldots, N$) as $\mathbf{H}_j(t)$ and stack them in the herder population position vector $\mathbf{H}:=\left[\mathbf{H}_1, \dots, \mathbf{H}_N\right]$.
Herders are modeled as velocity-saturated single integrators--similarly to the target population under the same kinematic assumptions--subject to the same obstacle repulsion:
\begin{equation}
\label{eq:herder_dynamics}
\dot{\mathbf{H}}_j \;=\; \left\{ \mathbf{u}_j \;+\; \mathbf{F}_\mathrm{o}(\mathbf{H}_j)\right\}_{v_\mathrm{H}},
\end{equation}
where \(\mathbf{u}_j \in [-v_\mathrm{H}, v_\mathrm{H}]^2 \) is the control input, and $\mathbf{F}_\mathrm{o}(\mathbf{q})$ is the short-range repulsion from obstacles as in \eqref{eq:Fc_def}. We remark that this potential force merely acts as a collision-avoidance mechanism, ensuring that both herders and targets do not collide with obstacles. For this reason, \( \mathbf{F}_\mathrm{o} \) needs to be complemented by the control input $\mathbf{u}_j$ to be designed. 
Finally, $v_\mathrm{H}$ is the maximum herder velocity component.

\subsection{Control goal}
The goal is to design a control law for the \(N\) herders such that they guide the \(M\) targets into the goal region \( \Omega_\mathrm{G} \), while avoiding the obstacles present in the environment. 
We consider a decentralized, communication-free setting. Moreover, we assume unknown models of the agents when designing control policies.

Following ~\cite{napolitano2025hierarchical}, we measure the fraction of targets inside the goal at time \(t\):
\begin{equation}
\label{eq:chi_def}
\chi(t) \;=\; \frac{1}{M}\,\Big|\big\{\, i \in \{1,\dots,M\} : \mathbf{T}_i(t) \in \Omega_\mathrm{G} \,\big\}\Big|,
\end{equation}
where $\lvert \mathcal{A} \rvert$ denotes the cardinality of set $\mathcal{A}$.

Each herder \(j\) can sense an observation vector containing the positions of all the agents and the center of its closest obstacle $P_{\hat{j}}$. If multiple obstacles are at the same distance, one is randomly selected to define $P_{\hat{j}}$.
\begin{comment}

\begin{equation}
\mathbf{S}_j(t) \;=\; \Big[\, \{\mathbf{H}_k(t)\}_{k \in \mathcal{N}_j^{H}},\; \{\mathbf{T}_i(t)\}_{i \in \mathcal{N}_j^{T}},\; \mathbf{P} \,\Big] \in \mathbb{R},
\end{equation}
where \( \mathcal{N}_j^{H} \) and \( \mathcal{N}_j^{T} \) denote, respectively, the sets of herders and targets sensed by agent \(j\) (under full sensing they coincide with all indices).
\end{comment}

\begin{equation}
    \mathbf{S}_j(t)=\left[\mathbf{H}, \mathbf{T}, {\mathbf{P}}_{\hat{j}}\right] \in \mathbb{R}^{2(N+M+1)}
\end{equation}

The control input of herder \(j\) is sampled from a (possibly stochastic) decentralized policy conditioned on its observation:
\begin{equation}
\mathbf{u}_j(t) \;\sim\; \pi\!\left(\,\cdot \,\middle|\, \mathbf{S}_j(t) \right), 
\qquad j=1,\dots,N .
\end{equation}

The control objective is achieved if
\begin{equation}
\exists \;t_\mathrm{g} < +\infty \quad  \text{s.t.} \quad \chi(t_\mathrm{g}) \ge \chi^\star,
\end{equation}
with \( \chi^\star \in (0,1] \) being the desired minimum fraction of targets within \( \Omega_\mathrm{G} \) (e.g., \( \chi^\star = 1 \)).

\subsection{Shepherding Performance Metrics}
\label{sec:metrics}
To evaluate how effectively a candidate policy satisfies the above objective, we introduce the following metrics \cite{auletta2022}, where we consider a value of \( \chi^\star = 1 \):

\begin{itemize}
  \item \textit{Gathering time} \( t_\mathrm{g} \). First time instant in which the success fraction reaches \( \chi^\star \):
  \begin{equation}
    t_\mathrm{g} \;=\; \inf_{t \ge 0} \left\{\, t : \chi(t) \ge \chi^\star \,\right\}.
  \end{equation}

  \item \textit{Average Path Length} \( d(t) \). Mean distance travelled by each herder in the interval \( [0, t] \):
  \begin{equation}
    d(t) \;=\; \frac{1}{N} \sum_{i=1}^{N} \int_{0}^{t} \big\| \dot{\mathbf{H}}_i(\tau) \big\|_2 \, d\tau.
  \end{equation}
  Notice that, considering a negligible $\mathbf{F}_\mathrm{o}(\mathbf{H}_j(t))$ for almost every time instant in a simulation, this metric also serves as a proxy for the average control effort.
\end{itemize}

\section{Control Architecture}
\label{sec:control_architecture}
To address the complexity of the task, we extend the two-layer hierarchical control architecture previously introduced for learning-based shepherding in obstacle-free environments~\cite{napolitano2025hierarchical}, adapting it to cluttered scenarios with obstacles.

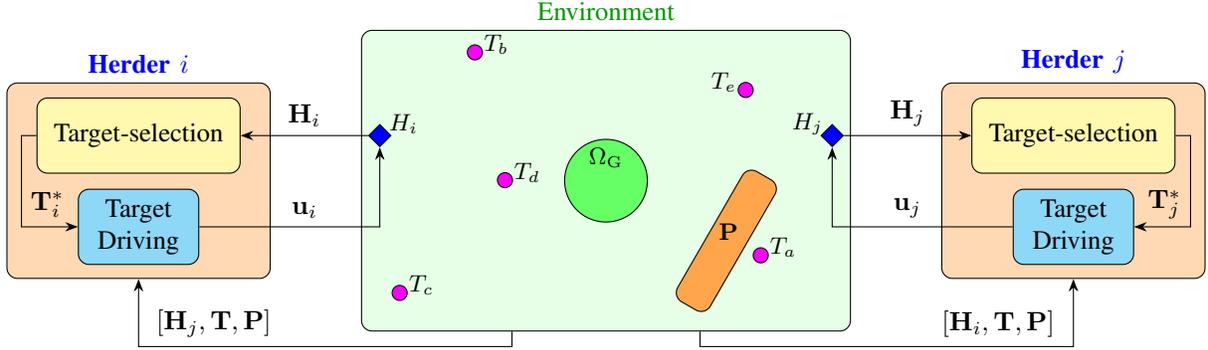
\begin{figure*}[ht]
    \centering
    \vspace{0.1cm}
    \begin{tikzpicture}[node distance=1cm, auto, >=Stealth]

        % Define custom colors
        \definecolor{h_blue}{RGB}{0, 0, 255}
        \definecolor{t_magenta}{RGB}{255, 0, 255}
        \definecolor{g_green}{RGB}{102, 255, 102}

        % --- Environment Block (rectangular and larger) ---
        \node[rectangle, draw, rounded corners, align=center, fill=green!10,
              minimum width=6.5cm, minimum height=4cm,
              label={[green!60!black, above, yshift=-0cm]Environment}] (env) {};

        % --- Obstacle (orange rectangle with rounded corners and manual orientation) ---
        % orientato ortogonalmente all’asse che collega il centro del riquadro al centro della goal
        \node[rectangle, draw, fill=orange!70, rounded corners=4pt,
      minimum width=2cm, minimum height=0.5cm, rotate=60, label={[right, font=\small]$\mathbf{P}$}]
      at ($(env.center) + (1.6,-0.8)$) (obstacle) {};

        % --- Green Circle (Goal Region, centrata) ---
        \node[circle, draw, fill=g_green, minimum size=1.1cm, inner sep=0pt,
              label={[below, font=\small]$\Omega_\mathrm{G}$}]
              at (env.center) {};

        % --- Herders (blue diamonds), moved higher and aligned with target-selection policy ---
        \node[diamond, draw, fill=h_blue, minimum size=0.3cm, inner sep=0pt,
        label={[right, font=\small]$H_i$}]
        at ([shift={(0.25,0.60)}]env.west) (herder_i) {};

        \node[diamond, draw, fill=h_blue, minimum size=0.3cm, inner sep=0pt,
        label={[left, font=\small]$H_j$}]
        at ([shift={(-0.25,0.60)}]env.east) (herder_j) {};

        % --- Targets (magenta circles) ---
        \node[circle, draw, fill=t_magenta, minimum size=0.2cm, inner sep=0pt,
              label={[right, font=\small]$T_a$}]
              at ([shift={(-1.2,-3)}]env.north east) {};
        \node[circle, draw, fill=t_magenta, minimum size=0.2cm, inner sep=0pt,
              label={[right, font=\small]$T_b$}]
              at ([shift={(-5,-0.3)}]env.north east) {};
        \node[circle, draw, fill=t_magenta, minimum size=0.2cm, inner sep=0pt,
              label={[right, font=\small]$T_c$}]
              at ([shift={(-6,-3.5)}]env.north east) {};
        \node[circle, draw, fill=t_magenta, minimum size=0.2cm, inner sep=0pt,
              label={[right, font=\small]$T_d$}]
              at ([shift={(-4.6,-2)}]env.north east) {};
        \node[circle, draw, fill=t_magenta, minimum size=0.2cm, inner sep=0pt,
              label={[left, font=\small]$T_e$}]
              at ([shift={(-1.4,-0.8)}]env.north east) {};

        % --- Agent i block ---
        \node[rectangle, draw, rounded corners, align=center, fill=orange!30,
              minimum width=3.5cm, minimum height=2.6cm, left=1.2cm of env,
              label={[h_blue, above]\textbf{Herder ${i}$}}] (agent_i) {};

        % Target-selection policy (yellow)
        \node[rectangle, draw, rounded corners, fill=yellow!40,
              align=center, minimum width=2.7cm, minimum height=1cm]
              at ($(agent_i) + (0, 0.6)$) (selection_i)
              {Target-selection};

        % Driving policy (cyan)
        \node[rectangle, draw, rounded corners, fill=cyan!40,
              align=center, minimum width=1.6cm, minimum height=1cm,
              below=0.2cm of selection_i] (driving_i) {Target\\Driving};

        % --- Agent j block ---
        \node[rectangle, draw, rounded corners, align=center, fill=orange!30,
              minimum width=3.5cm, minimum height=2.6cm, right=1.2cm of env,
              label={[h_blue, above]\textbf{Herder ${j}$}}] (agent_j) {};

        % Target-selection policy (yellow)
        \node[rectangle, draw, rounded corners, fill=yellow!40,
              align=center, minimum width=2.7cm, minimum height=1cm]
              at ($(agent_j) + (0, 0.6)$) (selection_j)
              {Target-selection};

        % Driving policy (cyan)
        \node[rectangle, draw, rounded corners, fill=cyan!40,
              align=center, minimum width=1.6cm, minimum height=1cm,
              below=0.2cm of selection_j] (driving_j) {Target\\Driving};

        % --- Arrows and feedback connections ---
        \draw[->] (selection_i.west) -| ($(selection_i.west) - (0.2,0)$)
            |- node[right, yshift=0.3cm] {$\mathbf{T}_i^*$} (driving_i.west);

        \draw[->] (driving_i.east) node[above, xshift=1.4cm] {$\mathbf{u}_i$}
            -| (herder_i.south);

        \draw[->] ($(env.south) - (1.25,0)$) -| ++(0,-0.2)
            -| node[above, xshift=1cm] {$[\mathbf{H}_j, \mathbf{T}, \mathbf{P}]$} (agent_i.south);

        \draw[->] (herder_i.west) -- node[above, xshift=0cm] {$\mathbf{H}_i$} (selection_i.east);

        \draw[->] (selection_j.east) -| ($(selection_j.east) + (0.2,0)$)
            |- node[left, yshift=0.3cm] {$\mathbf{T}_j^*$} (driving_j.east);

        \draw[->] (driving_j.west) node[above, xshift=-1.4cm] {$\mathbf{u}_j$}
            -| (herder_j.south);

        \draw[->] ($(env.south) + (1.25,0)$) -| ++(0,-0.2)
            -| node[above, xshift=-1cm] {$[\mathbf{H}_i, \mathbf{T}, \mathbf{P}]$} (agent_j.south);

        \draw[->] (herder_j.east) -- node[above] {$\mathbf{H}_j$} (selection_j.west);

    \end{tikzpicture}

    \vspace{0.3cm}
    \caption{Two-layer hierarchical feedback control scheme based on reinforcement learning, adapted from ~\cite{napolitano2025hierarchical}. Each herder $H_{i,j}$ detects the positions of other agents and determines the target $T_{i,j}^*$ via a \textit{target-selection heuristic}. The corresponding motion is then governed by the \textit{driving} policy, which outputs the velocity command $\mathbf{u}$ of the herder.}
    \label{fig:control_architecture}
\end{figure*}

As illustrated in Fig.~\ref{fig:control_architecture}, the overall control problem is decomposed into two subtasks: \textit{target selection} and \textit{target driving}. At each time step, every herder queries the corresponding modules to determine its next action.

The high-level decision module selects which target to influence, following the heuristic strategy proposed in~\cite{lama2024}. The low-level control module, designed via reinforcement learning, computes the herder’s velocity to drive the selected target toward the goal based on local observations.

We first train the low-level policy in a compact one-target/one-herder state–action space (Sec.\ref{sec:1h1t}). The resulting single-agent controller is then fixed and extended to the multi-agent case through a heuristic high-level controller (Sec.\ref{sec:multiagent_scenario}), which, despite being suboptimal, avoids the curse of dimensionality. This hierarchical decomposition alleviates the burden of multi-agent decision making while promoting specialization, modularity, and scalability.

\section{Learning and Deployment}
\subsection{Learning the Low-Level Driving Policy}
\label{sec:1h1t}
We begin by training the low-level driving policy in a nominal environment comprising one herder, one target, and a single obstacle. The policy is parameterized by a neural network and trained using the actor–critic Proximal Policy Optimization (PPO) algorithm~\cite{schulman2017}, suitable for continuous action spaces. In this scenario, we learn the policy \(\pi(\cdot,\mathbf{S}(t))\), where the local observation vector is \(\mathbf{S}(t) = [\mathbf{H}, \mathbf{T}, \mathbf{P}] \in \mathbb{R}^6\). This compact observation vector keeps the input dimension low, allowing for a lightweight network architecture. The controller outputs a continuous action \(\mathbf{u}(t) \in [-v_\mathrm H,v_\mathrm H]^2\).
Both the actor and critic networks take \(\mathbf{S}(t)\) as input and consist of five hidden layers with 64 ReLU units each. The critic produces a scalar value estimate, while the actor outputs the mean of a Gaussian distribution over actions, with trainable standard deviations. During validation, the standard deviation is set to zero for deterministic execution. 

The reward function encourages target approach, goal-directed guidance, and parsimonious control effort~\cite{covone_hierarchical_2025}:
\begin{equation}
\label{eq:reward_driving}
\begin{split}
r_\mathrm{D} = & -k_a \| \mathbf{T}(t_k) - \mathbf{H}(t_k)\|_2 \mathbb{I}_{\Omega_0 \backslash \Omega_\mathrm{G}} (\mathbf{T}(t_k)) \\
& -k_s (\|\mathbf{T}(t_k) \|_2 - \rho_\mathrm{G} ) \mathbb{I}_{\Omega_0 \backslash \Omega_\mathrm{G}}(\mathbf{T}(t_k)) \\
& -k_c \| \mathbf{u}(t_k)\|_2,
\end{split}
\end{equation}
where \(\mathbb{I}_\mathcal{A}(\mathbf{q})\) is the indicator function of set \(\mathcal{A}\), equal to 1 if \(\mathbf{q}\in\mathcal{A}\) and 0 otherwise.  
The first term keeps the herder close to the target, the second pushes the target toward the goal region \(\Omega_\mathrm{G}\), and the third penalizes large control actions. The gains are chosen such that \(k_s > k_a > k_c\).

To promote obstacle-aware behavior, we employ a simple curriculum on the initial conditions. With probability \(p_{\text{obs}}\), the target is initialized uniformly at random within a conical region \(\Omega_0^* \subset \Omega_0\) located behind the obstacle relative to the goal (see Fig.~\ref{fig:reward_with_inset}, inset). With probability \(1 - p_{\text{obs}}\), both agents are initialized uniformly in \(\Omega_0 \setminus \Omega_0^*\).  
This setup encourages the emergence of avoidance strategies while enabling the policy to generalize across both obstacle-free and cluttered scenarios.

\subsection{Decentralized Multi-Agent Deployment}
\label{sec:multiagent_scenario}
Building upon the low-level policy learned in Sec.\ref{sec:1h1t}, we deploy multiple herders guided by the decentralized target-selection heuristic module described in Sec.\ref{sec:control_architecture}. In so doing, the high-level controller applies the rule in~\cite{lama2024} to select targets and supply inputs to the fixed low-level driving module.

At each decision step, a herder \( {H}_j \) retains only the targets for which it is the closest herder. This nearest-agent exclusion avoids redundant assignments and induces implicit coordination among herders without communication. Formally, a target \( {T}_i \) is admissible for \( {H}_j \) if the following condition is satisfied:
\begin{equation}
\| \mathbf{T}_i - \mathbf{H}_j \|_2 \;<\; \| \mathbf{T}_i - \mathbf{H}_a \|_2 \quad \text{for all } j \neq a.
\end{equation}
Among the admissible set of targets, the herder selects the one that is the farthest from the goal, identified with the origin, so as to reduce the worst-case distance to \( \Omega_\mathrm{G} \) and promote global progress. If no admissible target is available, the herder moves at a speed $v_\mathrm{H}$ toward the boundary of the goal region and re-engages as soon as a valid assignment arises.

This rule is local and scalable, accommodates the lack of target cohesion, and provides structured role allocation while keeping the low-level controller unchanged across varying populations sizes.

\section{Numerical Validation}
\label{sec:validation}
We evaluate the proposed control strategies in two settings: (i) a nominal single-herder/single-target (1H–1T) scenario with one obstacle, and (ii) a multi-agent configuration with 10 herders, 100 targets, and three obstacles. Simulation details follow~\cite{lama2024} and are reported in Appendix~\ref{appendix:numerical}.

For each simulation, $\mathbf{P}_k$ is uniformly sampled within a circular region of radius $R > \rho_G$ centered at the origin.
Given the fixed obstacle configuration, the initial positions of both herders and targets are uniformly sampled within $\Omega_0$.

\subsection{Single-Herder/Single-Target (1H–1T)}
\begin{figure}[t] 
    \centering 
    \begin{overpic}[width=\columnwidth]{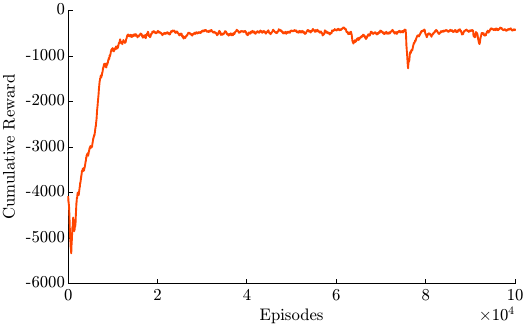} 
    \put(60,10){\includegraphics[width=0.37\linewidth]{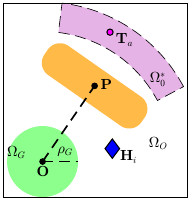}} 
    \end{overpic} 
    \caption{Cumulative reward during PPO training in the 1H--1T setup with a single rectangular obstacle. The inset illustrates the initialization region \( \Omega_0^* \subset \Omega_0 \) behind the obstacle with respect to the goal region, where the target is placed with probability \( p_{\text{obs}} = 0.5 \) to promote obstacle-aware behavior. The final cumulative reward maintains a nearly constant value with limited fluctuations, suggesting that the agent has converged to a stable policy capable of effectively completing the task in most training scenarios.} 
    \label{fig:reward_with_inset} 
\end{figure}

Training was conducted over \(10^5\) episodes with \(p_{\text{obs}}=0.5\).
Figure~\ref{fig:reward_with_inset} shows the evolution of the cumulative reward during PPO training, which reaches a plateau with small fluctuations in less than about $2\times 10^4$ episodes, indicating convergence to a stable policy capable of completing the task. %in most training scenarios.

The learned low-level driving policy was then validated against a vortex-based heuristic strategy that handles obstacle avoidance via tangential force fields~\cite{medio}. 
We performed \(E=1000\) validation episodes with random uniform initial conditions and \(p_{\text{obs}}=0.5\).
Both approaches proved robust, with the heuristic approach achieving a \(96.5\%\) success rate and the PPO-based policy reaching \(99.3\%\).

\begin{figure}[t]
    \centering

    % === Prima riga: metriche ===
    \begin{subfigure}[b]{0.48\columnwidth}
        \centering
        \includegraphics[width=\linewidth]{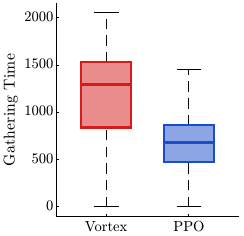}
        \caption{}
        \label{fig:metric_a}
    \end{subfigure}
    \hfill
    \begin{subfigure}[b]{0.48\columnwidth}
        \centering
        \includegraphics[width=\linewidth]{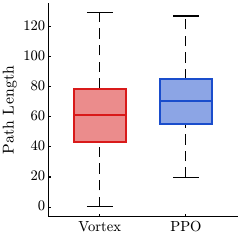}
        \caption{}
        \label{fig:metric_b}
    \end{subfigure}

    \vspace{0.3cm}

    % === Seconda riga: traiettorie ===
    \begin{subfigure}[b]{0.48\columnwidth}
        \centering
        \includegraphics[width=\linewidth]{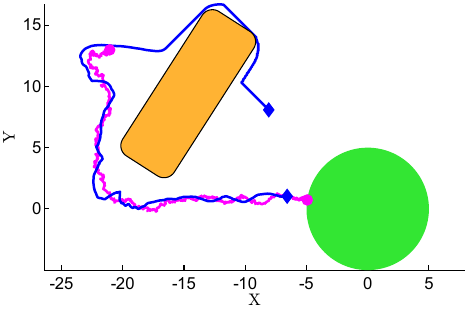}
        \caption{}
        \label{fig:metric_c}
    \end{subfigure}
    \hfill
    \begin{subfigure}[b]{0.48\columnwidth}
        \centering
        \includegraphics[width=\linewidth]{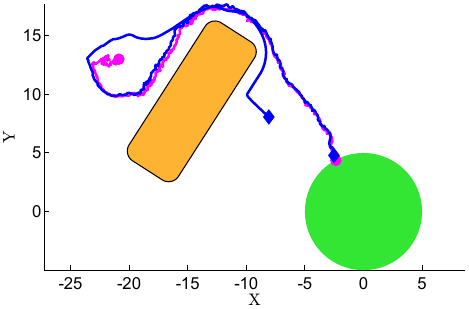}
        \caption{}
        \label{fig:metric_d}
    \end{subfigure}

    \caption{Comparison between the vortex heuristic and the PPO-based strategy. 
    Subfigures (a)–(b) show the gathering time and path length metrics; 
    subfigures (c)–(d) illustrate the corresponding obstacle-avoidance trajectories for the vortex and PPO strategies, respectively.}
    \label{fig:metric_comparison}
\end{figure}

As shown in Fig.~\ref{fig:metric_comparison}(a)–(b), the learning-based policy consistently achieves shorter gathering times \(t_{\mathrm{g}}\) with comparable average path lengths. 
This improvement stems from the different placement strategies used by the two strategies. The vortex controller maintains a fixed standoff from the target, whereas the PPO policy adapts its position to optimize guidance and reduce \(t_{\mathrm{g}}\). 
Figures~\ref{fig:metric_comparison}(c)–(d) further illustrate obstacle-avoidance behaviors under identical initial conditions. While the vortex strategy follows obstacle boundaries, the PPO-based policy produces smoother, more efficient trajectories and shorter settling times.

Overall, the learned policy demonstrates stable performance and outperforms the heuristic approach, benefiting from its optimization-based design.

\subsection{Multi-Agent: 10 Herders, 100 Targets (10H–100T)}
We evaluate a multi-agent scenario with \(N=10\) herders and \(M=100\) targets in an environment containing \(C=3\) rectangular obstacles whose positions are randomly and uniformly generated. We perform \(E=1000\) validation episodes with random uniform initial conditions and \(p_{\text{obs}}=0.5\).

Each herder executes the same low-level policy learned in the single-agent setup, assuming it can sense only the closest obstacle. High-level coordination follows the decentralized target-assignment heuristic strategy described in Sec.~\ref{sec:multiagent_scenario}. 

The proposed strategy achieves a success rate of \(99.7\%\), demonstrating both robustness and generalization of the learned control policy. The policy exhibits strong scalability and coordination capabilities, with an average gathering time of \(9.49 \pm 3.38 \times 10^{3}\) a.u. and a mean path length per herder of \(2.43 \pm 7.56 \times 10^{2}\) a.u., confirming consistent performance despite environmental complexity. 

\begin{figure}[t]
    \centering
    \begin{overpic}[width=\columnwidth, trim=0 20 0 0, clip] {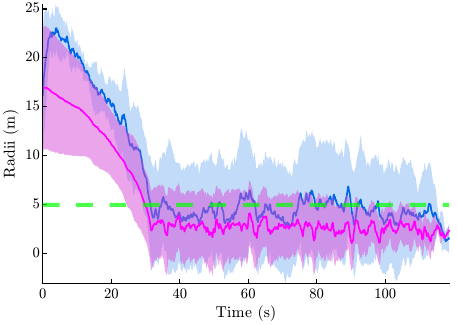}
        \put(60,33){\includegraphics[width=0.29\linewidth]{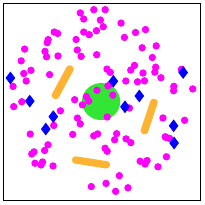}}
    \end{overpic}
    \caption{Evolution of the mean and standard deviation of target (magenta) and herder (blue) distances from the goal center during a representative 10H--100T episode with three rectangular obstacles. All target radii eventually fall below the goal threshold \( \rho_\mathrm{G} = 5 \) (green dashed line), confirming effective gathering. Herders subsequently enter the goal region as containment is not modeled. The inset shows the initial configuration of targets, herders, obstacles, and the goal region.}
    \label{fig:radii_targets_herders}
\end{figure}

Figure~\ref{fig:radii_targets_herders} shows the time evolution of the mean and standard deviation of the radial distances of targets and herders from the goal center during a representative episode. Initially, both populations are uniformly distributed within the initialization region \(\Omega_0\); as the episode progresses, the herders collectively drive the targets toward the goal region \(\Omega_\mathrm{G}\). The mean target radius decreases until all targets fall below the goal threshold \(\rho_\mathrm{G}=5\), indicating successful gathering despite the presence of multiple obstacles in the environment. These results highlight the scalability of the proposed control strategy to complex multi-agent environments without requiring retraining of the learned policy.

\paragraph{Computational effort}
All experiments ran on \texttt{Macbook Air M3} with a memory of \texttt{$24$\,GB}. Training the low-level policy for \(10^5\) episodes (\(\approx\)$1.2 \times 10^8$ environment steps) required approximately \(3\) hours with an average throughput of  approximately \(1.11\times10^4\) steps/s per PPO update. At deployment, single-step inference costs \(0.451 \pm 0.321\) ms per herder.

\section{Conclusions}
We addressed the problem of achieving multi-agent shepherding of non-cohesive targets in cluttered environments via a hierarchical, learning-based solution. Our key contributions include (i) a low-level PPO \emph{driving} policy trained in a minimal one-herder/one-target/one-obstacle setting that \emph{generalizes effectively} to multi-agent deployments; (ii) an architecture coupling this low-level policy with a decentralized target-selection heuristic strategy that preserves modularity and scalability; and (iii) convincing numerical evidence of collision-free guidance with improved efficiency over a vortex-based heuristic method from previous literature. Together, these elements deliver a scalable, modular, and model-free framework for sparse, indirect control in complex domains.

Future work will focus on integrating explicit \emph{containment} behaviors together with a \emph{learning-based target-selection} layer in the spirit of~\cite{napolitano2025hierarchical}, strengthening safety guarantees via control-theoretic tools (e.g., control barrier functions~\cite{ames2019control} and related approaches~\cite{hamandi}) and validating the approach on physical platforms with realistic sensing, actuation, and uncertainty. Another important direction is to relax sensing assumptions by removing knowledge of the obstacle-center position, enabling herders to operate using only local/perceptual cues (e.g., boundary proximity) rather than relying on any global geometric information.

%%%%%%%%%%%%%%%%%%%%%%%%%%%%%%%%%%%%%%%%%%%%%%%%%%%%%%%%%%%%%%%%%%%%%%%%%%%%%%%%
\section*{Appendix}
\label{appendix:numerical}
Our learning framework is implemented in Python~3.9, using Gymnasium for reinforcement-learning environments and PyTorch for neural network policies. The PPO algorithm follows the implementation in~\cite{implementazioneppo}, with hyperparameters initialized from~\cite{schulman2017} and subsequently refined to improve sample efficiency and training stability. The parameters used in the simulations reported in Sec.~VI, together with the PPO hyperparameters, are summarized in Tables~\ref{table_params} and~\ref{table_ppo}, respectively.
The code and the videos of the reported experiments are publicly available at 
\href{https://github.com/SINCROgroup/Decentralized-Shepherding-of-Non-Cohesive-Swarms-Through-Cluttered-Environments-via-DRL}{https://tinyurl.com/rl-shepherding}.

\begin{table}[h]
\caption{Simulation parameters derived from ~\cite{lama2024}. The first-order model is numerically integrated through the Euler–Maruyama method with a time step of $\Delta t = 0.01$~s.}
\label{table_params}
\begin{center}
\begin{tabular}{|c||c|c||c|}
\hline
Parameter & Value & Parameter & Value \\
\hline
$\rho_\mathrm{G}$ & 5 & $\beta$ & 3 \\
$R$ & 25 & $D$ & 0.1 \\
$L$ & 10 & $\lambda$ & 2.5 \\
$S$ & 1 & $d_\mathrm{o}^*$ & 3  \\
$v_\mathrm{H}$ & 8 & $v_\mathrm{T}$ & 7.5  \\
\hline
\end{tabular}
\end{center}
\end{table}

\begin{table}[h]
\caption{Hyperparameters and reward coefficients employed during the PPO training process.}
\label{table_ppo}
\begin{center}
\begin{tabular}{|c||c|c||c|}
\hline
Hyperparameter & Value & Hyperparameter & Value \\
\hline
Learning rate & 5e-4 & Rollout length & 4096 steps \\
Discount factor & 0.98 & Parallel environments & 8 \\
GAE parameter & 0.95 & Epochs per update & 10 \\
Clipping parameter & 0.2 & Hidden layers & 5 \\
Entropy coefficient & 0.01 & Neurons per layer & 64 \\
Value loss weight & 0.5 & $k_a$ & 5e-2 \\
Gradient clipping & 0.5 & $k_s$ & 1e-1 \\
Minibatches per update & 128 & $k_c$ & 2e-2 \\
\hline
\end{tabular}
\end{center}
\end{table}


%%%%%%%%%%%%%%%%%%%%%%%%%%%%%%%%%%%%%%%%%%%%%%%%%%%%%%%%%%%%%%%%%%%%%%%%%%%%%%%%

\bibliographystyle{IEEEtran}
\bibliography{refs}

\end{document}